\documentclass[twocolumn,aps,prl,superscriptaddress]{revtex4-2}
\usepackage{amssymb} 
\usepackage{amsmath,bm} 
\usepackage{graphicx} 
\usepackage[normalem]{ulem} 
\usepackage{multirow} 
\usepackage[colorlinks,linkcolor=blue,urlcolor=blue,citecolor=blue]{hyperref} 
\usepackage{lipsum} 
\usepackage[usenames, dvipsnames]{xcolor} 
\usepackage{tensor} 
\usepackage{isotope} 
\usepackage{amsmath} 
\usepackage{paracol}
\allowdisplaybreaks[4] 

\renewcommand{\sout}{\bgroup \color{red} \ULdepth=-.5ex \ULset}
   
\begin{document}
\title{Unfolding Baryon Number Fluctuations from Correlations of Light Nuclei Production in Heavy-Ion Collisions}

\author{Yi-Heng Feng}
%\email{23110190019@m.fudan.edu.cn}
\affiliation{Department of Physics and Center for Field Theory and Particle Physics, Fudan University, Shanghai 200438, China}
\affiliation{Shanghai Research Center for Theoretical Nuclear Physics, NSFC and Fudan University, Shanghai 200438, China}

\author{Che Ming Ko} 
\email{ko@comp.tamu.edu} 
\affiliation{Cyclotron Institute and Department of Physics and Astronomy, Texas A\&M University, College Station, Texas 77843, USA}

\author{Xiaofeng Luo} 
\email{xfluo@mail.ccnu.edu.cn} 
\affiliation{Institute of Particle Physics and Key Laboratory of Quark and Lepton Physics (MOE), Central China Normal University, Wuhan, 430079, China}

\author{Yu-Gang Ma}
\email{mayugang@fudan.edu.cn}
\affiliation{Shanghai Research Center for Theoretical Nuclear Physics, NSFC and Fudan University, Shanghai 200438, China}
\affiliation{Key Laboratory of Nuclear Physics and Ion-beam Application (MOE), Institute of Modern Physics, Fudan University, Shanghai 200433, China}

\author{Kai-Jia Sun}
\email{kjsun@fudan.edu.cn}
\affiliation{Shanghai Research Center for Theoretical Nuclear Physics, NSFC and Fudan University, Shanghai 200438, China}
\affiliation{Key Laboratory of Nuclear Physics and Ion-beam Application (MOE), Institute of Modern Physics, Fudan University, Shanghai 200433, China}

\author{Song Zhang}
\email{song\_zhang@fudan.edu.cn}
\affiliation{Shanghai Research Center for Theoretical Nuclear Physics, NSFC and Fudan University, Shanghai 200438, China}
\affiliation{Key Laboratory of Nuclear Physics and Ion-beam Application (MOE), Institute of Modern Physics, Fudan University, Shanghai 200433, China}

\date{\today}
\begin{abstract}
Event-by-event fluctuations of the baryon number, which is mostly carried by protons and neutrons, in relativistic heavy-ion collisions provide a sensitive probe for locating the conjectured critical point in the quantum chromodynamics (QCD) phase diagram. Since current experiments have limited access to neutron fluctuations because detectors are largely insensitive to neutrons, measurements of (net-)proton fluctuations are often used as a proxy for (net-)baryon number fluctuations. Although direct measurements of neutron fluctuations are challenging, their information are encoded in the production and correlations of light nuclei, when they are formed through   coalescence  of nucleons at kinetic freeze-out. Here, we propose to unfold neutron fluctuations from correlations among light nuclei produced in heavy-ion collisions. Model calculations validate this approach and show that baryon number fluctuations can be  unfolded up to the third order. For fourth and higher-order cumulants, however, the uncertainties become sizable, indicating that further methodological developments and refinements are required.
\end{abstract}

\maketitle

\emph{Introduction.}{\bf ---}
Various experimental observations have indicated that a novel state of matter consisting of deconfined quarks and gluons, namely the quark-gluon plasma (QGP), is   created in ultra-relativistic nuclear collisions~\cite{ALICE:2022wpn, STAR:2005gfr, Gyulassy:2004zy, Asakawa:2000wh, Rischke:1995mt}. Quantum Chromodynamics (QCD), the theory of strong interactions among quarks and gluons, suggests that the phase transition between the QGP and the hadron gas is a smooth crossover at vanishing baryon density, while effective theories suggest there is a strong first-order phase transition at high baryon densities~\cite{Aoki:2006we, Bhattacharya:2014ara, deForcrand:2009zkb, Li:2011ee, deForcrand:2014tha}. A critical end point (CEP) of the second-order phase transition is expected to exist between these two regions.   Searching for the possible signatures of the CEP and the first-order phase transition is one of the main goals for ongoing and future heavy-ion experiments~\cite{Pandav:2022xxx, STAR:2022vlo, CBM:2016kpk, Shiltsev:2019rfl, Xiaohong:2018weu, Braun-Munzinger:2014lba}.

A key feature of the CEP is the divergence of correlation length, which is expected to induce large density fluctuations of conserved charges~\cite{Stephanov:1999zu, Hatta:2003wn,  Stephanov:2008qz, Athanasiou:2010kw, Stephanov:2011pb, Shao:2019xpj}. Among these observables, event-by-event (net-)baryon number fluctuations have garnered particular interest because they are more sensitive to the CEP than electric charge and strangeness~\cite{Hatta:2003wn, Stephanov:1999zu, Luo:2015doi, Luo:2017faz,Chen:2024aom}. The STAR collaboration has recently observed a non-monotonic energy dependence of net-proton number fluctuations in the Beam Energy Scan (BES) program~\cite{STAR:2020tga,STAR:2025zdq}. The latest measurements reveal a deviation from model predictions at a maximum significance of up to 5$\sigma$~\cite{ STAR:2025zdq}. Meanwhile, the measurements have been extended to higher-order cumulants and lower energies~\cite{STAR:2022vlo, STAR:2022etb}. In addition to event-by-event baryon number fluctuations, intermittency index and light nuclei yield ratios have also been proposed as a probe for the critical point as they are sensitive to the baryon density fluctuations~\cite{Sun:2017xrx, Sun:2018jhg, Wu:2019mqq}. It is suggested that large nucleon density fluctuations near the critical point would lead to an enhancement of light nuclei yield ratios $N_t N_p/N^2_d$, and possible signatures have been observed by the STAR Collaboration~\cite{STAR:2022hbp}.

The baryons produced in heavy-ion collisions are predominately protons and neutrons, especially at intermediate and lower collision energies. However, experiments so far have only measured proton fluctuations, since neutrons are electrically neutral and difficult to measure at RHIC and the LHC. Proton fluctuations are therefore often taken as a proxy under the assumption that proton and neutron fluctuations are independent and identical.   On the other hand, theoretical studies~\cite{Vovchenko:2021yen, Vovchenko:2021kxx, Vovchenko:2020kwg} show that baryon number fluctuations are systematically deviating from proton fluctuations, indicating that protons and neutrons are not statistically independent. Thus, a proper determination of baryon number fluctuations requires information of neutron number fluctuations and neutron-proton correlations.

In this study, we extract neutron-number fluctuations from correlations among light nuclei (e.g., deuterons and tritons) and then unfold the event-by-event baryon-number fluctuations. The method is based on the idea that nucleon number fluctuations and correlations are encoded in the production of light nuclei formed via e.g. nucleon coalescence at kinetic freeze-out. We derive analytical expressions that connect proton and neutron number fluctuations to those of light nuclei, and validate these relations using a multi-phase transport (AMPT) model~\cite{Lin:2004en, Lin:2021mdn}, which incorporates the full dynamics of both partonic and hadronic phases. Our results show that baryon number fluctuations can be reliably unfolded up to third order, whereas the extraction of fourth- and higher-order cumulants has appreciable uncertainties.

\emph{Unfolding baryon number fluctuations from correlations of light nuclei.}{\bf ---} 
The moments of baryon number distribution can be expanded in terms of the moments of proton number distribution and correlations between proton and neutron numbers, and the results are summarized as follows,
\begin{equation}
\begin{aligned}
    &\langle (\delta N)^2 \rangle 
    = 2  C_2(N_p) + 2\langle \delta N_p \delta N_n \rangle ,\\
    &\langle (\delta N)^3 \rangle  
    = 2 C_3(N_p) + 6 \langle (\delta N_p)^2 \delta N_n \rangle ,\\     
    &\langle (\delta N)^4 \rangle  = 2\langle (\delta N_p)^4 \rangle + 8\langle (\delta N_p)^3 \delta N_n \rangle  \\ & \qquad \qquad \, \, 
     +6 \langle (\delta N_n)^2 (\delta N_p)^2 \rangle.
     \label{eq:Cn}
\end{aligned}
\end{equation}
Here $\delta X = X - \langle X \rangle$ with $X$ denoting the number of baryon ($N$), proton ($N_p$), or neutron ($N_n$), and $C_n(X)$ denotes the $n-$th order cumulant. In obtaining Eq.~(\ref{eq:Cn}), we have assumed that the moments of neutrons are identical to those of protons. If protons and neutrons are uncorrelated, the baryon-number moments are simply twice those of protons. Previous analyses of baryon number fluctuations often neglected these correlations—either setting them to zero or approximating the proton–neutron covariance using the second moment of the proton distribution~\cite{Hatta:2003wn, Stephanov:2008qz, Kitazawa:2012at}. As demonstrated below, such approximations are inadequate  because of  non-negligible  neutron-proton correlations.  

Since proton and neutron are the  two isospin states of a  nucleon, each nucleon can be treated as a proton with probability \(r\) and as a neutron with probability \((1-r)\). For an event class containing \(N^i\) nucleons per event, the proton number \(N_p^i\) in that class follows a binomial distribution \(\mathrm{Bino}(r, N^i)\), while the neutron number \(N_n^i\) follows \(\mathrm{Bino}(1-r, N^i)\):  
\begin{equation}
    \begin{aligned}
        P(N_p^i) &= \binom{N^i}{N_p^i} r^{N_p^i}(1-r)^{\,N^i-N_p^i}, \\
        P(N_n^i) &= \binom{N^i}{N_n^i} (1-r)^{\,N_n^i} r^{\,N^i-N_n^i}.
    \end{aligned}
\end{equation}
In high-energy collisions, the  probability \(r\) is approximately \(0.5\), indicating that protons and neutrons follow statistically identical distributions, i.e., \(C_{n}(N_p) = C_{n}(N_n)\).  
Because the total nucleon number \(N\) fluctuates from event to event, the unconditional variance of the proton number must be evaluated using the law of total variance: the expectation of the conditional variance plus the variance of the conditional expectation. Accordingly, we obtain  
\begin{equation}
    \frac{C_2}{C_1}(N_p)  
    = \frac{\langle C_2(N_p \mid N)\rangle_N + C_2[\langle N_p \mid N \rangle]}{\langle \langle N_p \mid N \rangle \rangle_N}  
    = 1 - \frac{\alpha}{2},
    \label{derive_C2C1p}
\end{equation}
where the first term, \(\langle C_2(N_p \mid N)\rangle_N\), represents the average conditional variance of the proton number \(N_p\) for a fixed nucleon number \(N\), averaged over all possible values of \(N\). In statistical terms, it quantifies the typical spread in proton numbers within each subset of events sharing the same nucleon number, followed by an overall averaging across all \(N\).  The second term, \(C_2[\langle N_p \mid N \rangle]\), denotes the variance of the conditional expectation of \(N_p\) with respect to \(N.\) Conceptually, it measures how the mean proton number (at fixed \(N\)) varies as \(N\) itself fluctuates between events.  Here, \(\alpha = 1-\frac{C_2}{C_1}(N)\) characterizes the deviation of the baryon-number cumulant ratio from unity, independent of the specific distribution of \(N\).  
The unconditional covariance between protons and neutrons is given by  
\begin{equation}
    \mathrm{Cov}(N_n,N_p) = -\frac{\alpha \langle N \rangle}{4}.
    \label{derive_Covpn}
\end{equation} 
 
If the nucleon number \(N\) follows a Poisson distribution, then \(\frac{C_2}{C_1}(N) = 1\). In this limit, the proton number also follows a Poisson distribution and protons and neutrons are statistically independent.  Previous studies indicate that \(\alpha\) is positive; from Eq.~(\ref{derive_Covpn}), this implies an intrinsic negative correlation between protons and neutrons.   It is therefore essential to treat proton–neutron correlations properly, yet such information is not directly accessible in current experiments. To address this limitation, we extract the required correlation information from light nuclei. 

Three primary mechanisms for the production of light nuclei are currently discussed in the literature. The first is the statistical model~\cite{Andronic:2010qu, Vovchenko:2019pjl, Andronic:2016nof, Andronic:2012dm}, which assumes that light nuclei are emitted together with hadrons from the quark–gluon plasma (QGP) at chemical freeze-out. The second is the coalescence model~\cite{Scheibl:1998tk, Sun:2018mqq, Zhao:2018lyf, Oh:2009gx, Feng:2024tmc, Bellini:2020cbj}, in which light nuclei are formed through the coalescence of nucleons at kinetic freeze-out  of  the hadronic phase. The third is the transport approach~\cite{Sun:2022xjr, Wang:2023gta}, which describes light-nucleus production   from multi-body scatterings  during the hadronic stage of heavy ion collisions. Within coalescence and transport models, where light nuclei are formed from nucleons in the hadronic phase rather than emitted directly from the QGP, these nuclei naturally carry information about neutron degrees of freedom. Consequently, combining proton-fluctuation measurements with those of light nuclei provides an indirect yet powerful means to infer neutron fluctuations

We begin with deuteron production in heavy-ion collisions.  Assuming that each proton–neutron pair has a probability \(q\) of forming a deuteron, the expected deuteron yield is \(\langle N_p N_n q \rangle\).  Imposing strict baryon number conservation, we can then evaluate the covariance between protons and deuterons as
\begin{equation}
\begin{aligned}
    {\rm Cov}(N_p, N_d) &= \langle N_d (N_p - N_d) \rangle - \langle N_d \rangle \langle N_p - N_d \rangle \\
    &\approx \left[ C_2(N_p) + Cov(N_p, N_n) \right] \frac{\langle N_d \rangle}{\langle N_p \rangle} - C_2(N_d).
    \label{expand_Covpd}
\end{aligned}
\end{equation}

Several assumptions enter the derivation of Eq.~\eqref{expand_Covpd}.  First, in addition to the proton–neutron covariance, our calculation involves the second-order mixed correlation term \(\langle (\delta N_p)^2 \delta N_n \rangle\).  A mixed distribution has been shown to approximate the measured baryon distribution well~\cite{Bzdak:2018uhv}.  
Using this distribution, we find  
\begin{equation}
    \frac{\langle (\delta N_p)^2 \delta N_n \rangle}{\langle \delta N_p \delta N_n \rangle} 
    = \frac{1}{2} \frac{C_3(N) - C_2(N)}{C_2(N) - C_1(N)} 
    \approx \frac{C_2}{C_1}(N).
    \label{deltp2n1}
\end{equation}
Second, we assume a sufficiently large nucleon number, leading to \(N_p \gg 1\) and \(q \approx \langle N_d \rangle / \langle N_p \rangle^2\).  Third, we assume that deuteron production does not significantly alter the proton number distribution.  This is justified by experimental observations: in central Au+Au collisions at BES energies, the event averaged proton number at mid-rapidity is typically between 20 and 40, while the deuteron-to-proton ratio ranges from approximately 0.3\% to 2\%.  Finally, it is not necessary to assume that all nucleon pairs share an identical probability of forming a deuteron.  As long as this probability remains independent of the proton and neutron numbers, the result derived above remains valid.  

The above proton–deuteron covariance can be generalized to obtain higher-order correlations between protons and neutrons. Besides deuterons, the next lightest nuclei are \(^3\mathrm{He}\) and \(^3\mathrm{H}\) (both with mass number \(A=3\)).  Similarly, we assume that each nucleon combination has a probability \(q\) of forming \(^3\mathrm{He}\) or \(^3\mathrm{H}\).  Under this assumption, the expected yields are approximately \(\langle N_n \binom{N_p}{2} q \rangle \approx \langle N_n N_p^2 q / 2 \rangle\) for \(^3\mathrm{He}\), and \(\langle N_p \binom{N_n}{2} q \rangle \approx \langle N_p N_n^2 q / 2 \rangle\) for \(^3\mathrm{H}\).  

The corresponding covariances between protons and these nuclei are then given by
\begin{widetext}
\begin{equation}
    \mathrm{Cov}(N_p,N_{^{3}\mathrm{He}}) 
    \approx \frac{\langle N_{^{3}\mathrm{He}} \rangle}{\langle N_p \rangle} \Bigg[ \frac{C_3}{C_1}(N_p) + \frac{\mathrm{Cov}(N_p^3,N_n)}{\langle N_p \rangle^2} + 2\Big(C_2(N_p) - \mathrm{Cov}(N_p,N_n)\Big) \Bigg] - 2 C_2(N_{^{3}\mathrm{He}}),
    \label{expand_phe3}
\end{equation}    
\begin{equation}
    \mathrm{Cov}(N_p,N_t)
    \approx \frac{\langle N_t \rangle}{\langle N_p \rangle} \Bigg[ C_2(N_p) + \frac{C_2}{C_1}(N_p)^2 - 2\mathrm{Cov}(N_p,N_n) + \frac{\mathrm{Cov}(N_p^2,N_n^2)}{\langle N_p \rangle^2} \Bigg] - C_2(N_t),
    \label{expand_ptriton}
\end{equation}
\end{widetext}
where no additional assumptions beyond those used for deuterons are introduced.  These covariances involve higher-order powers of proton and neutron numbers.  

From Eqs.~\eqref{expand_Covpd}, \eqref{expand_phe3}, and \eqref{expand_ptriton}, we see that, except for the proton–neutron covariances, all other quantities are experimentally measurable.  This indicates that by measuring correlations between protons and light nuclei, together with their individual fluctuations, the proton–neutron covariances can be extracted.  These covariances correspond to the proton–neutron correlations appearing in Eq.~\eqref{eq:Cn}.  Using the approximation in Eq.~\eqref{deltp2n1}, the cumulants of the baryon number distribution -- commonly employed in event-by-event fluctuation analyses -- can be expressed in terms of these covariances and the cumulants of the proton number distribution:
\begin{equation}
\begin{aligned}
  &C_2(N) = 2C_2(N_p) + 2\mathrm{Cov}(N_p,N_n),\\ 
  &C_3(N) = 2C_3(N_p) + 6\mathrm{Cov}(N_p,N_n)\frac{C_2}{C_1}(N),\\  
  &C_4(N) = 2C_4(N_p) + 8\mathrm{Cov}(N_p^3,N_n) + 6 \mathrm{Cov}(N_p^2,N_n^2) \\   
  &\qquad - 60 \mathrm{Cov}(N_p,N_n)^2 - 48\langle N_p \rangle^2 \mathrm{Cov}(N_p,N_n) \\  
  &\qquad - 72 \mathrm{Cov}(N_p,N_n) C_2(N_p).
  \label{restroed_Cn}
\end{aligned}
\end{equation}
All terms on the right-hand side of Eq.~\eqref{restroed_Cn} are now experimentally accessible.  Consequently, baryon number fluctuations can be reconstructed from the measured fluctuations and correlations of protons and light nuclei. 

\begin{figure}[!h]
\centering
\includegraphics[width=8.3cm]{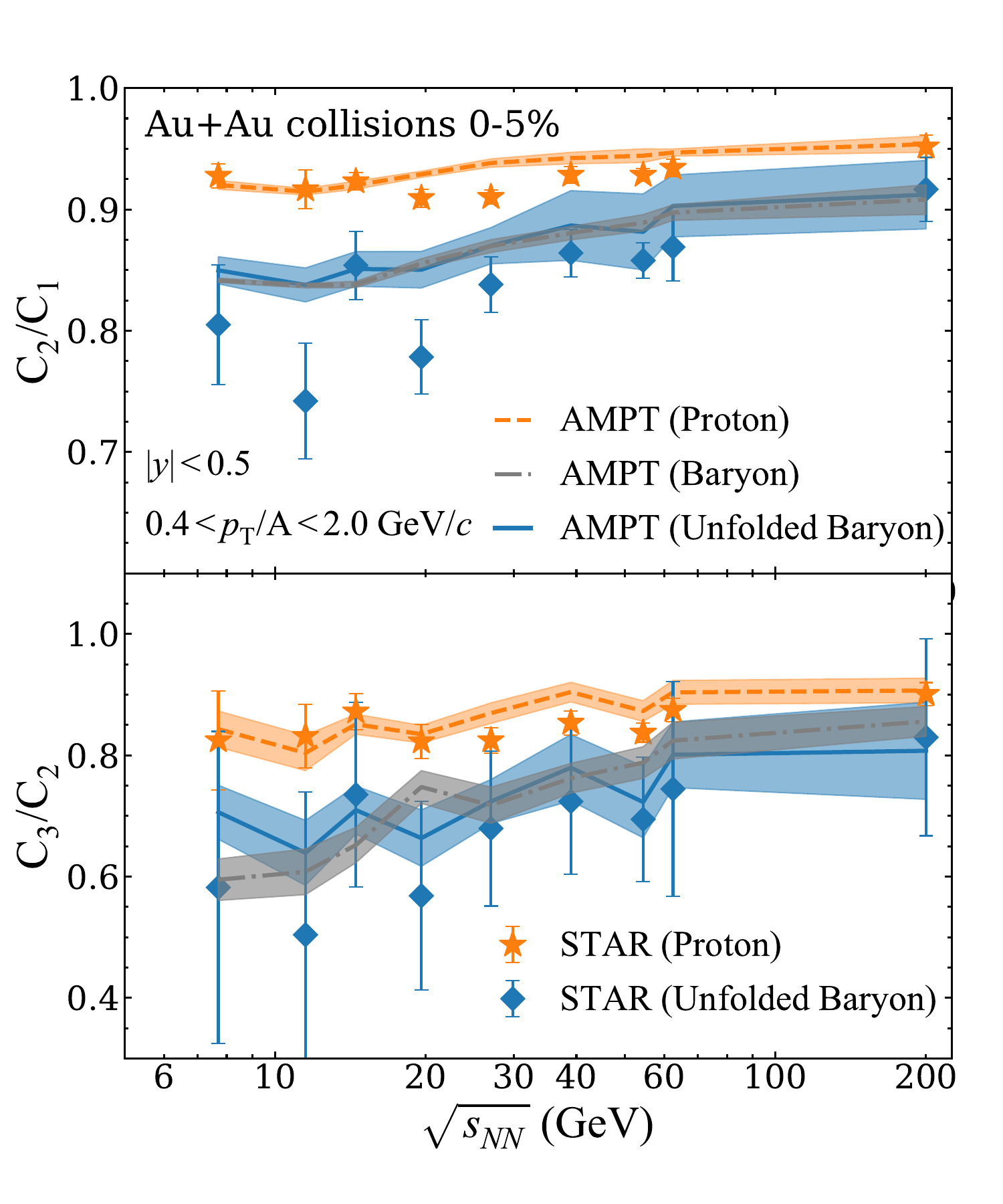}
\caption{Cumulant ratios of proton number and baryon number distributions as a function of collision energy $\sqrt{s_{NN}}$ in central Au+Au collisions. The upper panel shows $C_2 / C_1$, and the lower panel shows $C_3 / C_2$. The orange bands and stars represent the cumulant ratios of proton number distributions from the AMPT model and experiments. The grey bands represent the cumulant ratios of baryon number distributions from the AMPT model. The blue bands and diamonds represent the cumulant ratios of baryon number distributions restored by our method, both in the AMPT model and experiments. Error bands are obtained directly  from the statistical fluctuations in numerical calculations, while experimental data errors were computed from the data via the error propagation formula: $\frac{\sigma_{XY}}{\langle XY \rangle} \approx \sqrt{\left( \frac{\sigma_X}{\langle X \rangle} \right)^2 + \left( \frac{\sigma_Y}{\langle Y \rangle} \right)^2}$. The experimental data are taken from the STAR Collaboration~\cite{STAR:2021iop, STAR:2023ebz}. }
\label{pic: C2C1C3C2}
\end{figure}

\emph{Model validation.}{\bf ---} 
The accuracy of above method for unfolding baryon number fluctuations from light nuclei can be assessed  by employing a transport model that is capable of describing the full evolution of heavy-ion collisions. Specifically, we use the AMPT model~\cite{Lin:2004en, Wang:1991hta} to simulate collisions across the BES energy range of 7.7–200 GeV, adopting the default parameters for all energies, and then we generated $5\times 10^6$  Monte Carlo events for each energy.

We follow the procedure used by the STAR Collaboration for centrality classification and fluctuation calculations.  For the centrality classification, it is determined from the multiplicity of charged particles at mid-rapidity, excluding protons and light nuclei to avoid self-correlation effects~\cite{Luo:2017faz}.  For the 0–5\% most central collisions, we further subdivide the sample into 30 uniform \(N_{ch}\) bins to implement the centrality bin width correction (CBWC)~\cite{Luo:2011ts, Luo:2013bmi}.  

Since the AMPT model does not explicitly include light nuclei production, we incorporate a nucleon coalescence model on top of AMPT to calculate their yields at the kinetic freeze-out stage.  In the coalescence model, the formation probability of a nucleus of mass number $A$ is given by the value of its Wigner function~\cite{Chen:2006vc}, i.e.,
\begin{equation}
    P_A=g_c 8^{A - 1} \exp \left[ - \sum_{i = 1}^{A - 1} \left( \frac{q_{i}^{2}}{\sigma_{i}^{2}} + \sigma_{i}^{2} k_{i}^{2} \right) \right],
\end{equation}
where \(g_c\) is the spin degeneracy factor, \(q_i\) and \(k_i\) are the relative coordinates and momenta in the Jacobian coordinate system of the light nucleus center-of-mass frame~\cite{Mattiello:1996gq, Chen:2003tn}, and \(\sigma_i\) is the reduced mass.   For nucleons of equal mass, \(\sigma^2 = \frac{2A}{3(A-1)} \langle r_A^2 \rangle\), where \(\langle r_A^2 \rangle\) denotes the root-mean-square radius of the light nucleus.  
For deuterons, we use \(g_c = 3/4\) and \(\sigma =2.26~\mathrm{fm}\)~\cite{Ropke:2008qk, Sun:2017ooe}. 

To include baryon-number conservation effects, we compute the Wigner function for each nucleon pair and use it as the Monte Carlo probability that the pair coalesces into a deuteron.
If the sampling succeeds, a deuteron is added to the event with momentum equal to the total momentum of the nucleon pair. The constituent nucleons are then removed from the event to ensure they do not participate in the formation of other deuterons.  

Using the full particle information from the AMPT calculations, numerical evaluations of all physical quantities  needed in our method show that the negative correlation between protons and deuterons obtained numerically is stronger than the analytical estimate from Eq.~\eqref{expand_Covpd}.  Further investigation indicates that this discrepancy arises from the non-independence between the deuteron formation probability \(q\) and the proton number \(N_p\).  Although centrality is determined by the charged-particle multiplicity excluding protons, the source size and proton number remain correlated. This violates the prior assumption that \( q \) and \( N_p \) are statistically independent, as the coalescence probability \( q \) is inversely proportional to the source size, due to the larger phase-space distance between nucleons.  This introduces an additional, non-negligible correlation between protons and deuterons, which is difficult to incorporate analytically.  

To quantify this effect, we use the AMPT model, finding that it is most pronounced at \(\sqrt{s_{NN}} = 7.7\) GeV and nearly negligible at \(\sqrt{s_{NN}} = 200\) GeV.  We model the energy dependence using an inverse proportional function, scaled by the deuteron-to-proton ratio \(d/p\).  After fitting in AMPT, the impact is approximately \(- \frac{\langle N_d \rangle / \langle N_p \rangle}{\log(\frac{s}{\mathrm{GeV}^2})}\).  Accordingly, the Pearson coefficient between proton and deuteron numbers becomes
\begin{equation}
    \rho_{N_p,N_d} = \rho^{\text{0}}_{N_p,N_d} - \frac{\langle N_d \rangle / \langle N_p \rangle}{\log(\frac{s}{\mathrm{GeV}^2})},
\end{equation}
here \(\rho^{\text{0}}_{N_p,N_d}\) corresponds to the covariance in Eq.~\eqref{expand_Covpd}, which was derived under the assumption that proton number $N_p$ and coalescence probability $q$ are statistically independent ($\rho_{q,N_p} = 0$). The Pearson coefficient is related to the covariance via \(\rho_{X,Y} = \frac{\mathrm{Cov}(X,Y)}{\sqrt{C_2(X) C_2(Y)}}\).  
Thus, the proton–neutron covariance can be expressed in terms of the proton–deuteron covariance as
\begin{equation}
\begin{aligned}
    &\mathrm{Cov}(N_p,N_n) = \\ 
    &\left[ \left( \rho_{N_p,N_d} +  \frac{\langle N_d \rangle / \langle N_p \rangle}{\log (\frac{s}{\mathrm{GeV}^2})} \right) \sqrt{\frac{C_2(N_p)}{C_2(N_d)}}+1 \right] \frac{C_2}{C_1}(N_d)\langle N_p \rangle  \\ 
    &- C_2(N_p).
\end{aligned}
\label{restore_Covpn}
\end{equation} 

Substituting Eq.~\eqref{restore_Covpn} into Eq.~\eqref{restroed_Cn}, we can unfold the second- and third-order cumulants of the baryon number distribution from deuteron and proton fluctuations.  The results are shown in Fig.~\ref{pic: C2C1C3C2}, where the x-axis represents the collision energy \(\sqrt{s_{NN}}\) and the y-axis shows the cumulant ratios.  For a strict Poisson distribution, these ratios should equal 1.  Both experimental data and transport model calculations indicate that proton and baryon number distributions in the studied rapidity and momentum region deviate from a Poisson distribution.  

In Fig.~\ref{pic: C2C1C3C2}, orange bands represent cumulant ratios of the proton number distribution, while gray bands show baryon number cumulant ratios calculated in AMPT.  The baryon number cumulant ratios are systematically lower than the proton ones, consistent with MUSCI + SAM results~\cite{Vovchenko:2021kxx}.  Orange stars represent STAR Collaboration measurements~\cite{STAR:2021iop}, and AMPT reproduces the proton number fluctuations observed experimentally.  
From the upper panel, a significant deviation between the experimental \(\frac{C_2}{C_1}(p)\) and the AMPT result is observed near 20 GeV, particularly at 27 GeV, where the experimental measurement is 3\(\sigma\) lower.  The deviation in \(\frac{C_3}{C_2}(p)\), shown in the lower panel, is smaller but still noticeable. 
 
\begin{figure}[!t]
\centering
\includegraphics[width=9.0cm]{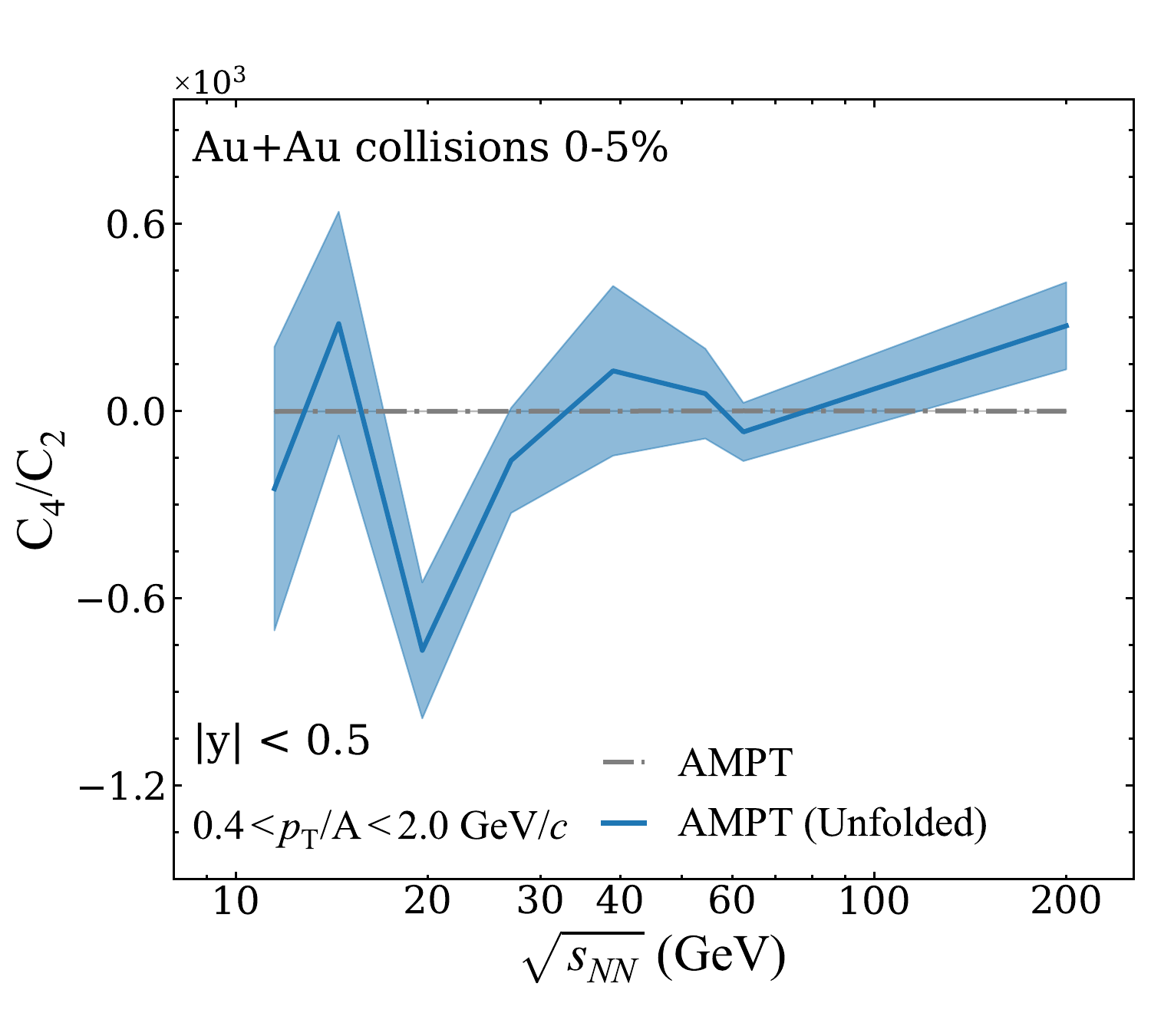}
\caption{The fourth-to-second cumulant ratio of baryon number distribution as a function of collision energy $\sqrt{s_{NN}}$ in the AMPT model. The gray band represents the result given by AMPT, while the blue band represents the result unfolded with light nuclei production.}
\label{pic: C4C2}
\end{figure}

The blue band and blue diamonds in the figure represent the baryon number cumulant ratios unfolded using our method.  
It is evident that this approach can accurately reconstruct baryon number fluctuations in the AMPT model, and the unfolded experimental results are in good agreement with the AMPT calculations.  Notably, the \(C_2 / C_1\) ratio of the baryon number shows deviations from the model at the two lowest collision energies, a feature not observed in the proton number distribution.  
In contrast, for the \(C_3 / C_2\) ratio of the baryon number, no significant deviation from theoretical predictions is seen, although large uncertainties may obscure potential signals.  Due to the lack of experimental data for triton and helium-3 fluctuations, we rely solely on AMPT to test the feasibility of reconstructing fourth-order baryon number cumulants using light nuclei.  As shown in Fig.~\ref{pic: C4C2}, the fourth-order cumulant ratio calculated in AMPT is  essentially unity.  However, because heavier light nuclei have lower yields and poorer statistics, the reconstructed results exhibit large uncertainties, often exceeding the magnitude of the values themselves.  Consequently, it remains challenging to draw definitive conclusions regarding the applicability of our method for fourth- and higher-order baryon number fluctuations. 

\emph{Isospin fluctuations.}{\bf ---}
The neutron-proton correlation is directly linked to isospin fluctuations. Their relation can be quantified as
\begin{equation}
    \delta I_3 = \frac{C_2(N_p - N_n)}{C_2(N_p + N_n)} \approx \frac{1-\rho_{N_p,N_n}}{1+\rho_{N_p,N_n}},
    \label{C2_isospin}
\end{equation}
where \(\rho_{N_p,N_n}\) denotes the Pearson coefficient between proton and neutron numbers.  Enhanced negative correlations correspond to increased isospin fluctuations.  

\begin{figure}[!t]
\centering
\includegraphics[width=9.0cm]{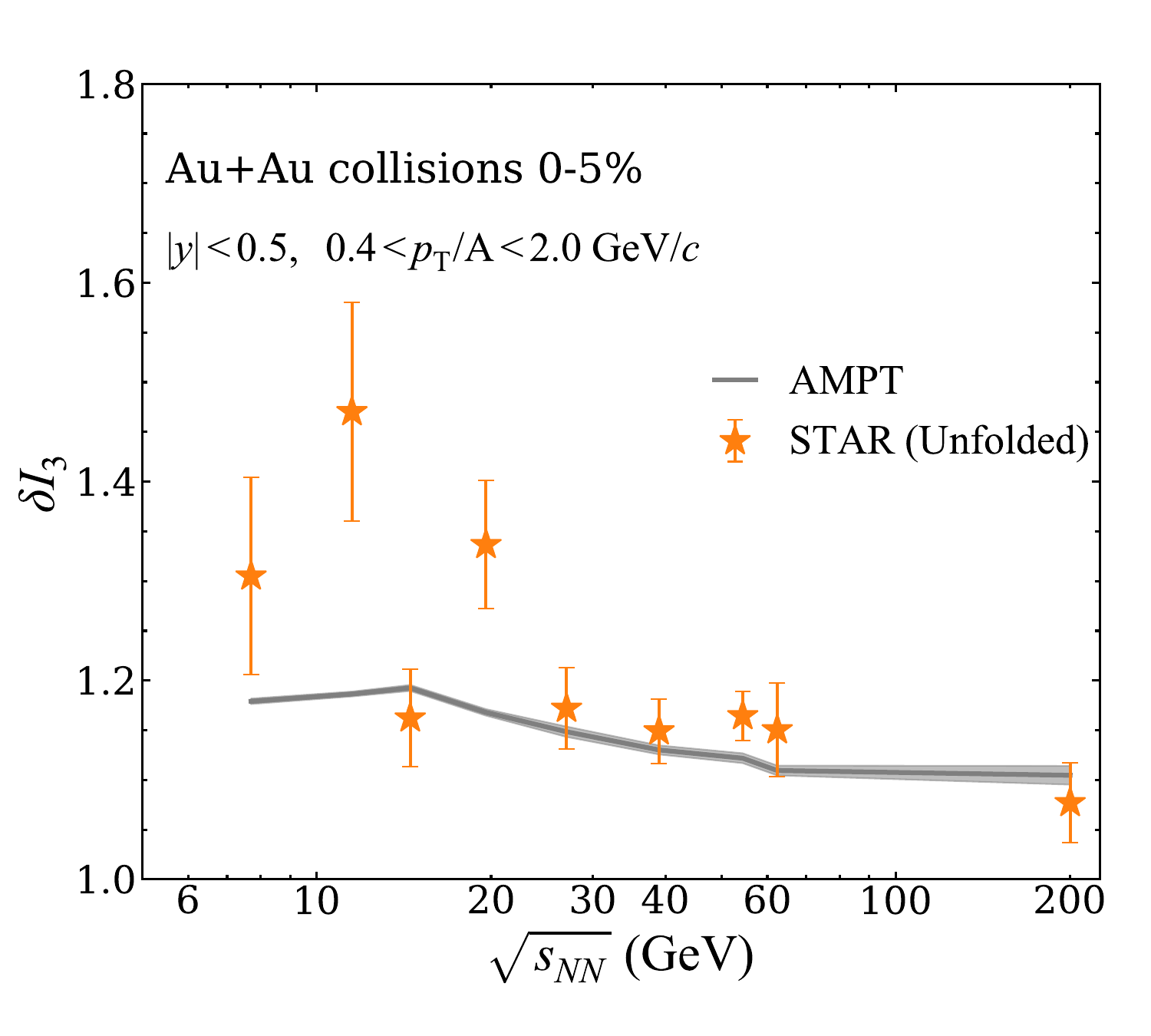}
\caption{Energy dependence of isospin fluctuation. The grey band is obtained in the AMPT model and orange stars represent the results unfolded from experimental data~\cite{STAR:2021iop, STAR:2023ebz}. The errors were calculated the same way as in Fig.~\ref{pic: C2C1C3C2}.}
\label{pic: c2isospin}
\end{figure}

Figure~\ref{pic: c2isospin} shows the isospin fluctuations calculated using Eq.~\eqref{C2_isospin} in both the AMPT model and existing experimental data.  The orange stars represent STAR measurements, while the gray band denotes the AMPT results.  We observe pronounced enhancement of isospin fluctuations at \(\sqrt{s_{NN}} = 11.4\) and 19.6 GeV, while at other energies, the AMPT model and experimental results are in good agreement.  

Unlike the relatively flat trend predicted by AMPT, the experimental energy dependence of isospin fluctuations is non-monotonic. The unfolded isospin fluctuations exhibit deviations from the AMPT model at $\sqrt{s_{NN}} = 11.5$~GeV and $\sqrt{s_{NN}} = 19.6$~GeV, reaching approximately $2.5\sigma$ at these two energies. This energy dependence seems to be a double-peak structure, which is anticipated by QCD theory~\cite{Sun:2018jhg}, with the first peak associated with the spinodal region~\cite{Steinheimer:2012gc} and the second with the critical end point (CEP)~\cite{Berdnikov:1999ph}. However, given the substantial uncertainties, we cannot confirm the reality of this structure. Further theoretical and experimental investigations are thus necessary. 

\emph{Conclusion.}{\bf ---} 
In the present study, we have proposed a method to extract information on neutron and baryon number fluctuations from measurements of correlations and fluctuations  of light nuclei produced in relativistic heavy-ion collisions. Such information is not directly accessible in current experiments. By deriving analytical relations that link baryon cumulants to proton cumulants and proton–neutron correlations, and by demonstrating how these correlations can be obtained from covariances involving deuterons, ${}^{3}\mathrm{He}$, and ${}^{3}\mathrm{H}$, we establish a framework that enables neutron fluctuations to be inferred from experimentally measurable observables.

Using AMPT simulations combined with a nucleon coalescence model, we  have shown that the proposed method reliably reconstructs the second- and third-order cumulants of the baryon number distribution. Applying this unfolding procedure to existing STAR data, we successfully extract $C_{2}$ and $C_{3}$ for central Au+Au collisions across a broad range of BES energies. The reconstructed baryon number cumulants exhibit deviations from transport-model expectations around $\sqrt{s_{NN}} \approx 20$ GeV, consistent with those observed in proton cumulants. Additional deviations, absent in proton-only measurements, appear near $\sqrt{s_{NN}} \approx 10$ GeV, reflecting enhanced proton–neutron correlations and increased isospin fluctuations.

Reconstructing the fourth-order cumulant requires event-by-event measurements of $^3$He and $^3$H, including their first and second cumulants and their covariances with proton number. Due to the low production rates of $A=3$ nuclei and the complexity of the higher-order expressions, the statistical uncertainties are too large to yield a reliable reconstruction of $C_{4}$. Since the fourth-order cumulant arises from subtracting two large moments, even small approximations can obscure the underlying physical signal. Future high-statistics measurements of light nuclei, together with improved modeling of their production~\cite{Wang:2023gta,ALICE:2025byl,Zhang:2025tfd} and machine learning techniques~\cite{Boehnlein:2021eym,He:2023zin}, will be essential for extending this approach to higher-order cumulants.

\emph{Acknowledgments.}{\bf ---}We thank Wen-Hao Zhou and Rui Wang for helpful discussions. This work was supported in part by the National Key Research and Development Project of China under Grant No. 2024YFA1612500, No. 2024YFA1610802, and No.2022YFA1604900;  the National Natural Science Foundation of China under contract No. 12422509, No. 12375121, No. 124B2102, No. 12147101, No. 12275054, No. 12025501, No. 12061141008, No. 12525509, and 12447102; the Natural Science Foundation of Shanghai under Grant No. 23JC1400200 and No. 23590780100; the Shanghai Pilot Program for Basic Research-Fudan University 21TQ1400100(22TQ006); the Guangdong Major Project of Basic and Applied Basic Research No. 2020B0301030008;  and the STCSM under Grant No. 23590780100. The computations in this research were performed using the CFFF platform of Fudan University.

%\bibliography{ref.bib} 

 
\end{document}